*Research article*

**Plant species accumulation curves are determined by evenness and spatial aggregation in drylands worldwide**

Niv DeMalach[1]* (nivdemalach@gmail.com)

Hugo Saiz[2]

Eli Zaady[3]

Fernando T. Maestre[2]

[1]Department of Ecology, Evolution and Behavior, The Hebrew University of Jerusalem, Givat Ram, Jerusalem, 91904, Israel

[2]Departamento de Biología y Geología, Física y Química Inorgánica, Universidad Rey Juan Carlos, c/ Tulipán s/n, 28933 Móstoles, Spain

[3]Department of Natural Resources, Institute of Plant Sciences, Agriculture Research Organization, Ministry of Agriculture, Gilat Research Center, Gilat 85280, Israel

* Corresponding author: Tel. 972-2-6584659, Fax: 972-2-6858711

**ABSTRACT**

Species accumulation curves (SAC), i.e. the relationship between species richness and the number of sampling units in a given community, can be used to describe diversity patterns while accounting for the well-known scale-dependence of species richness. Despite their value, the functional form and the parameters of SAC, as well as their determinants, have barely been investigated in plant communities, particularly in drylands. We characterized the SAC of perennial plant communities from 233 dryland ecosystems from six continents by comparing the fit of major functions (power-law, logarithmic and Michaelis-Menten). We tested the theoretical prediction that the effects of aridity and soil pH on SAC are mediated by vegetation attributes such as evenness, cover, and spatial aggregation. We found that the logarithmic relationship was the most common functional form, followed by Michaelis-Menten and power-law. Functional form was mainly determined by evenness while the SAC parameters (intercept and slope) were largely determined by spatial aggregation. In addition, aridity decreased small scale richness (intercept of SAC) but did not affect accumulation rate (slope of the SAC). Our results highlight the role that attributes such as spatial aggregation and evenness play as main mediators of the SAC of vegetation in drylands, the Earth´s largest biome.

1. **INTRODUCTION**

Understanding how biodiversity varies in space and time is one of the main scientific challenges of this century [1, 2]. Species richness, the number of species in a given area, is the simplest and most used biodiversity index [e.g. 3, 4, 5]. However, species richness is extremely sensitive to the spatial scale considered [6, 7], and this scale-dependence is a major source of divergence among studies [8-10]. The main solution for this scale-dependence is the use of "Species-Area Relationships" (hereafter SAR) when characterizing communities, which describe how the total number of species varies as the sampled area increases [11, 12]. Here we focused on a common SAR type known as '*Species Accumulation Curve*' (hereafter SAC), which describes the *expected* (mean) number of species as a function of the number of sampling units [13-15]. The expected richness is computed based on all possible combination of sampling units (disregarding their spatial location within a site). A main advantage of this computation approach is its ability to produce a smooth curve (rather than a step function). This curve could be described by simple mathematical functions [16]. Importantly, SAC is a specific type of SAR, since the size of the area sampled is a linear function of the number of sampling units ['SAR type III', 17].

Ecological communities could vary both in the type of function that best characterizes their SAC (hereafter 'Functional form') and in the parameter values of that function (hereafter 'SAC parameters'). The functional form of the SAC could be described by various functions [12] but the simplest ones are power law ('Arrhenius'), logarithmic ('Gleason') and Michaelis-Menten. These functions differ in their degree of 'saturation' (i.e. the speed at which the slope of the accumulation curve decreases), with Michaelis-Menten being the most 'saturated' function, Logarithmic function being intermediate and Power-law being the least 'saturated' (Fig. 1).

Ecological models propose that both functional form and SAC parameters are determined by the following proximate factors (hereafter *mediators*): density, evenness, spatial aggregation and species pool [13, 18, 19]. Both density (number of individuals per area) and evenness (similarly in relative abundance among species) increase richness at small spatial scales by increasing the probability of species detection [18, 19]. Nonetheless, their effect on richness declines with increasing the number of sampling units [13]. Spatial aggregation (intraspecific clustering in spatial distribution) decreases richness at small scales since aggregated species are less likely to be sampled, but this effect decreases as more sampling units are incorporated [13, 18, 19]. Species pool size (the number of species that could colonize a site as determined by evolutionary and historical processes) has a positive effect on richness at all scales although its relative importance should increase with increasing the number of sampling units[13].

A full understanding of SAC drivers includes a 'causal cascade' where abiotic factors affect the mediators (e.g. pH affects evenness), thereby affecting SAC functional form and parameters[13]. Recently, the effect of the different mediators on scale-dependent richness response to gradients was tested in animal communities [20]. In contrast, the effects of such mediators on plant SAC was almost never studied despite the evidences that plant SACs (and other SAR types) are affected by environmental gradients [9, 14, 21]. Here, we aimed to do so by using data gathered from 233 dryland sites from six continents [22, 23]. Drylands cover 41% of Earth's land surface and support over 38% of the human population [22]. These drylands are threatened by global land-use and climate changes that may further decrease water availability [24]. Previous studies with this database have revealed that abiotic factors such as aridity and pH are main determinants of diversity patterns at the site scale (30x30 $M^2$)[25, 26] . However, their effect on SAC has never been investigated. Hence, in this contribution we studied SAC patterns focusing on the following questions:

(1) What is the relative role of different mediators (evenness, density, spatial aggregation and species pool) in determining SAC patterns?
(2) How aridity and pH affect the SAC mediators? How these effects are translated into SAC patterns?

## 2. METHODS

### (a) Study area and fitting species accumulation curves

The study includes 233 sites representative of the major types of dryland vegetation from all continents except Antarctica, which cover a wide range of plant species richness (from 2 to 49) and environmental conditions (mean annual temperature and precipitation ranged from -1.8 to 28.2 °C, and from 66 to 1219 mm, respectively). In all sites, vascular perennial vegetation was sampled using a standardized protocol [22]. Each site included 80 sampling units (1.5x1.5m) located along four 30-m length parallel transects (20 sampling units per transect, eight meters distance between transects) where the presence and cover of each perennial species was estimated. For each site, a species accumulation curve (SAC) was built using the 'Vegan' R package [27]. Then, each SAC was fitted to the following functions:

(1) Power-law function: $S = b_0 \cdot A^{b_1}$

(2) Logarithmic function: $S = b_0 + b_1 \cdot \log(A)$

(3) Michaelis-Menten function: $S = \dfrac{b_0 \cdot A}{b_1 + A}$

In all functions, *S* is the number of species (the dependent variable), A is the number of sampling units (the independent variable) and $b_0$ and $b_1$ are the two (estimated) parameters. The best function for each site was chosen based on the lowest corrected Akaike Information Criterion (AICc) of the fitted model [28].

### (b) Mediators of species accumulation curve

We estimated potential mediators of SAC (spatial aggregation, evenness, density) using several indices. As an index for spatial aggregation, we calculated the slope of the relationship between the incidence (proportion of the sampling units where the species was found) and the log abundance of each species. The steeper the slope, the more aggregated the plant community [see 29, 30 for details]. For evenness we used Pielou's classical index [31]: $\frac{H}{H_{max}}$, where *H* is Shannon entropy index and $H_{max}$ is the Shannon value obtained for the most even community with the same number of species (i.e. a theoretical 'ideal community' where all species have equal abundance). Cover (sum of the relative cover of different species) was used as a proxy for density because we did not directly measured the number of individuals per area. Still, we separated between the total cover of woody species ('woody cover') and total cover of perennial herbs ('herbaceous cover') assuming the latter group may include more individuals for a given cover (more herbs than shrubs could grow in a given level of cover due to their typically smaller size).

The last mediator, species pool (the number of species that could colonize a site as determined by evolutionary and historical processes) could not be estimated in our observational dataset. Importantly, the common *proxy* for species pool, site richness (number of species found in a given site when all sampling units are combined) is inevitably determined by SAC, which may lead to a circular reasoning [32, 33]. Hence, we assessed the role of species pool only indirectly as the sum of all the effects of aridity and pH which are not mediated by the other mediators (see next sections for details).

### (c) Classification of the functional forms

A classification tree was used for testing whether the potential mediators (spatial aggregation, evenness, herbaceous cover and woody cover) were able to predict SAC functional form. The analysis was conducted using the R package 'party', which allow unbiased recursive partitioning based on conditional inference, thereby reducing the risk of overfitting [34]. Importantly, although in most sites the differences (ΔAICc) between the best model and the second best model were high (median ΔAICc =107), nine sites where the AICc differences were lower than seven (i.e. sites with high uncertainty regarding the best model) were excluded from this classification analysis [35].

### (d) Estimating the drivers of SAC parameters

We built a structural equation model (SEM) using the R package 'piecewiseSEM' that allows a flexible analysis based on the local estimation method [36]. The SEM estimated the causal effects of the potential mediators (spatial aggregation, evenness, herbaceous cover and woody cover) as well as the abiotic factors, aridity (1-evaporation/precipitation) and soil pH [37]. The model also included a 'direct' effect of aridity and pH on SAC parameters, which represent effects that are independent of the mediators. Since theory suggests that species pool is the only proximate factor (mediator) that can affect SAC besides spatial aggregation, evenness and density [13], any 'direct' effect of aridity and pH could be interpreted as a species pool mediated effect of these environmental factors.

Obviously, parameters of different SAC functional forms cannot be compared. Hence, we applied the logarithmic functional form for all sites since this function had the highest explanatory power across sites (the median $R^2$ for all sites was 0.99), meaning that even in cases where the model was not the 'best' in terms of AICc, it could still be used as an approximation. Nonetheless, 14 sites where the $R^2$ of logarithmic function was less than 0.90 were excluded from the analysis to avoid large biases. The logarithmic function includes two coefficients, an intercept (hereafter 'small-scale richness', $b_0$) and a slope (hereafter 'accumulation rate', $b_1$).

We transformed several variables to meet the assumptions of SEM: accumulation rate and aggregation were log transformed while all the bounded indices (evenness, aridity, herbaceous and woody cover) were logit transformed. In addition, and to avoid problems of spatial autocorrelation (independence among nearby sites) we used Moran Eigenvectors Maps that were built with the R package 'adespatial' [see 38]. The inclusion of these eigenvectors enables the reduction of potential bias in parameter estimation caused by unmeasured factors related to spatial autocorrelation such as disturbances, historical land-use or soil characteristics. For reducing these confounding effects as much as possible (i.e. applying the most conservative approach), we included all the 38 positive eigenvectors in all the relationships in the SEM. More details on the model formulation are found in the electronic supplementary material.

3. RESULTS

All the three functional forms were found in the drylands studied (Fig. 2). The logarithmic function was the most common (112 sites [48%]) followed by Michaelis-Menten (79 sites [34%]) and power-law (42 sites [18%]) functions. Evenness was the only variable that predicted SAC functional form in the classification tree, although its predictive power was relatively modest (Fig. 3). Power-law and logarithmic relationships were (similarly) common under low evenness levels. The logarithmic relationship was the most common form under intermediate levels of evenness, while Michalis-Menten was the most common form under high levels of evenness.

The results of the structural equation model (Fig 4) show the main determinants on SAC intercept ($b_0$, 'small scale richness') and slope ($b_1$, 'accumulation rate'). There was has a positive effect of aridity on woody cover and evenness, and a negative effect on herbaceous cover. Still, evenness had only a modest positive effect on small scale richness, and did not affect accumulation rate. Furthermore, there was no effect of cover on any of the SAC parameters. Aridity had no significant effects on spatial aggregation. In addition, there were negative effects of pH on evenness and on small-scale richness. Interestingly, spatial aggregation was found to be a main determinant of SAC parameters. Increasing aggregation reduced small-scale richness and increased accumulation rate. Importantly, the strongest effects of aridity and pH on SAC were 'direct' effects on small scale richness (i.e. an effect that was not mediated by cover, aggregation and evenness).

## 4. DISCUSSION

In this study we used, for the first time, species accumulation curves (SAC) for characterizing the richness of perennial plants in drylands worldwide. We found that SAC functional form is mainly determined by evenness and that SAC parameters are mainly determined by aggregation. In addition, we found that aridity decreases small scale richness ($b_0$, the intercept of the SAC) but does not affect accumulation rate ($b_1$, the slope of the SAC).

Debates regarding SAR functional forms date back to the beginning of the 20$^{th}$ century [12]. So far, the most common approach for describing SAR is based on the power-law function [12, 39]. While theoretical models suggest that SAR should become more 'saturated' (lower second derivative) as evenness increases [13, 19], a recent review of SAR functions found these suggestions unsupported by empirical data [12]. Our findings are fully consisted with theoretical models [13, 19]. The most even communities were characterized mostly the asymptotic Michaelis-Menten SAC. Communities with intermediate levels of evenness were mostly characterized by the logarithmic function that tends to 'saturate' faster than power law. Communities with high evenness were characterized by power law or logarithmic functions in similar proportion. Still, we found that evenness cannot fully predict the SAC functional form suggesting that there are possible important unmeasured factors (e.g. species pool). Small species pool of drylands (which could be lead to asymptotic relationship) may explain the low proportion of power-law SAC found in our study (18%). Many previous vegetation SAR were documented in tropical forests with larger species pool [39] which could increase the proportion of the power law functional form. In addition, most previous studies investigated other type of SAR rather than SAC [39]. Hence, it remained to be tested whether the differences between our results and previous studies reflect differences between drylands vegetation and other systems or differences between SAC and other SAR types.

Interestingly, in most communities (c. 96%) the different functional forms were very distinguishable (AICc > 7). In contrast with other SAR types (where different functions are not always distinguishable [12, 39]), SACs are smooth functions with large 'sample size' (sample size equals the amount of sampling unit) allowing high statistical power. However, despite the detectable differences between functional forms we found that the logarithmic function could be used as approximation for all SAC since its bias is relativity small as indicated by the high explanatory power ($R^2 > 0.90$ for 94% of the sites).

The structural equation model did not support our expectation that the effects of aridity and pH on SAC parameters should be mediated by aggregation, cover and evenness. Although aridity decreased herbaceous cover, there were no effects of cover on SAC parameters. A possible explanation for this finding is that woody and herbaceous covers are not good indicators of density in this dataset. In addition, evenness increased with aridity and decreased with increasing soil pH but the positive effect of evenness on small scale richness was modest. We found that SAC parameters were mostly affected by our spatial aggregation index [29]. Aggregation decreased small scale richness but increased accumulation rate. Again, this finding is in accordance with theory but has rarely been supported by empirical data [19]. Furthermore, our approach underestimates the role of aggregation since our index quantifies only aggregation at the sampling unit level. Nevertheless, it is possible that aggregation occurring across other scales could also affect SAC parameters. Surprisingly, while aggregation is an important mediator of SAC parameters, it was not affected by aridity or pH. Hence, the drivers of aggregations remain to be tested in future studies.

We found a 'direct' negative effect of aridity on small scale richness and a modest positive effect of pH on small scale richness. These effects may represent unmeasured factors such as historical and evolutionary processes leading to small species pool in more arid and more acidic sites. Unfortunately, this interpretation could not be rigorously tested using our dataset due to the difficulty in defining the species pool [32, 33]. Additionally, underestimation of the roles aggregation and density (see above) may also lead to overestimation of a 'direct' effect of aridity and pH [40].

**CONCLUSIONS**

Our findings support untested theoretical predictions [13, 18, 19] and question the automatic use of power law functions to describe SARs. The findings that evenness and spatial aggregation are main mediators of SAC highlight the role of understanding their main drivers (e.g. competition, heterogeneity, dispersal). Hence, we suggest that future theoretical and empirical studies will focus on the mechanism of spatial aggregation and evenness that determine diversity in several scales instead of focusing on species number in a given (arbitrary) scale.

The link between functional form and evenness has important implications for conservation as well as for any efforts to characterize the number of species in a given area. While it is reasonable to expect an asymptote (when enough sampling units are sampled) in even communities that tend to 'saturate', uneven communities are unlikely to reach an asymptote. Thus, uneven communities inevitably require extrapolation methods for estimating the total number of species [41]. In addition, the important role of aggregation in determining SAC slope highlights the bias obtained from simple extrapolation from small scales (often used for experiments) to large scales (the target of conservation efforts). Such extrapolation will underestimate the number of species in large areas where species show high aggregated spatial dispersion while overestimating species number in communities with low spatial aggregation.

## DATA ACCESSIBILITY

Data is available on figshare: https://figshare.com/s/b4010a4331d73d99239f

## AUTHORS' CONTRIBUTIONS

F.T.M. designed the field surveys and N.D. designed the data analysis. N.D. and H.S. analyzed the data. The first draft was written by N.D and all authors contributed substantially to revisions.

## COMPETING INTERESTS

We declare we have no competing interests.

## FUNDING

This work was funded by the European Research Council under the European Community's Seventh Framework Program (FP7/2007-2013)/ERC Grant agreement 242658 (BIOCOM). FTM and HS are supported by the European Research Council (ERC Grant agreement 647038 [BIODESERT]). HS is supported by a Juan de la Cierva-Formación grant from MINECO.

## ACKNOWLEDGMENT

We thank all the members of the EPES-BIOCOM network for the collection of field data, all the members of the Maestre lab for their help with data organization and management and Ron Milo for comments on previous versions of this manuscript.

**FIGURE CAPTIONS**

**Figure 1.**

Three examples of 'species accumulation curves' (SAC) characterized by different functional forms. (*power-law relationship* - 'Morata_CCA' [Spain], *logarithmic relationship* - 'Site A' [China], *Michaelis-Menten relationship* - 'Aspe' [Spain]). The black lines are the SAC (the mean number of species based on of all combination of sampling units), while the gray areas represent their 95% confidence intervals. The colored lines represent the trend line of the best fitted functional form (ΔAICc>200 in all sites). The upper panels show the SAC in a linear space. In this scale all functions are concave but Michalelis-Menten is the most 'saturated' (i.e. characterized by the fastest decrease in the slope) while power-law is the least 'saturated'. The lower panels show the same SAC (from the same sites) on a semi-logarithmic space (the x-axis is logarithmic). In this scale power-law function is convex, logarithmic function is linear and Michaelis-Menten function is concave.

**Figure 2.**

Pie charts describing the proportion of SAC functional forms in drylands from across the globe (sites are aggregated by country). Red – Power-law function, blue – logarithmic function, green – Michaelis-Menten function. Circle size represents the number of sites per country.

**Figure 3.**

A classification tree predicting the functional form of the species accumulation curve (A - Power-law, G - Logarithmic, M – Michaelis-Menten) based on Pielou's evenness index (other possible mediators where found to be non-significant). N=224 (nine sites were the ΔAICc was lower than seven were excluded from the analysis)

**Figure 4**.

Results of a structural equation model (SEM) of factors determining 'small scale richness' and 'accumulation rate' (i.e. the intercept and the slope of the species accumulation curve assuming a logarithmic relationship). Rectangles represent observed variables; unidirectional arrows represent significant (P<0.05) positive (solid lines) and negative causal effects (dashed lines). The numbers above the unidirectional arrows are standardized coefficients (effect size). The bidirectional grey arrows represent positive (solid line) and negative (dashed line) correlations (the numbers near the lines are Pearson correlation coefficients). The $R^2$ values for the endogenous variables are, 0.37 (pH), 0.24 (Woody cover), 0.33 (Evenness), 0.35 (Herbaceous cover), 0.86 (Small-scale richness) and 0.79 (Accumulation rate). The model allows for non-causal correlations ('Woody cover ~ Herbaceous cover, Accumulation rate ~ Small-scale richness, Evenness ~ Herbaceous cover, Woody cover ~ Evenness, Woody cover ~

Aggregation). The model is in agreement with the data (local estimation, AICc =-1581, Fisher.C=6.99, p-value=0.14). All the parameters of the model are summarized in table S1. N=219

**FIGURES**

**Figure 1.**

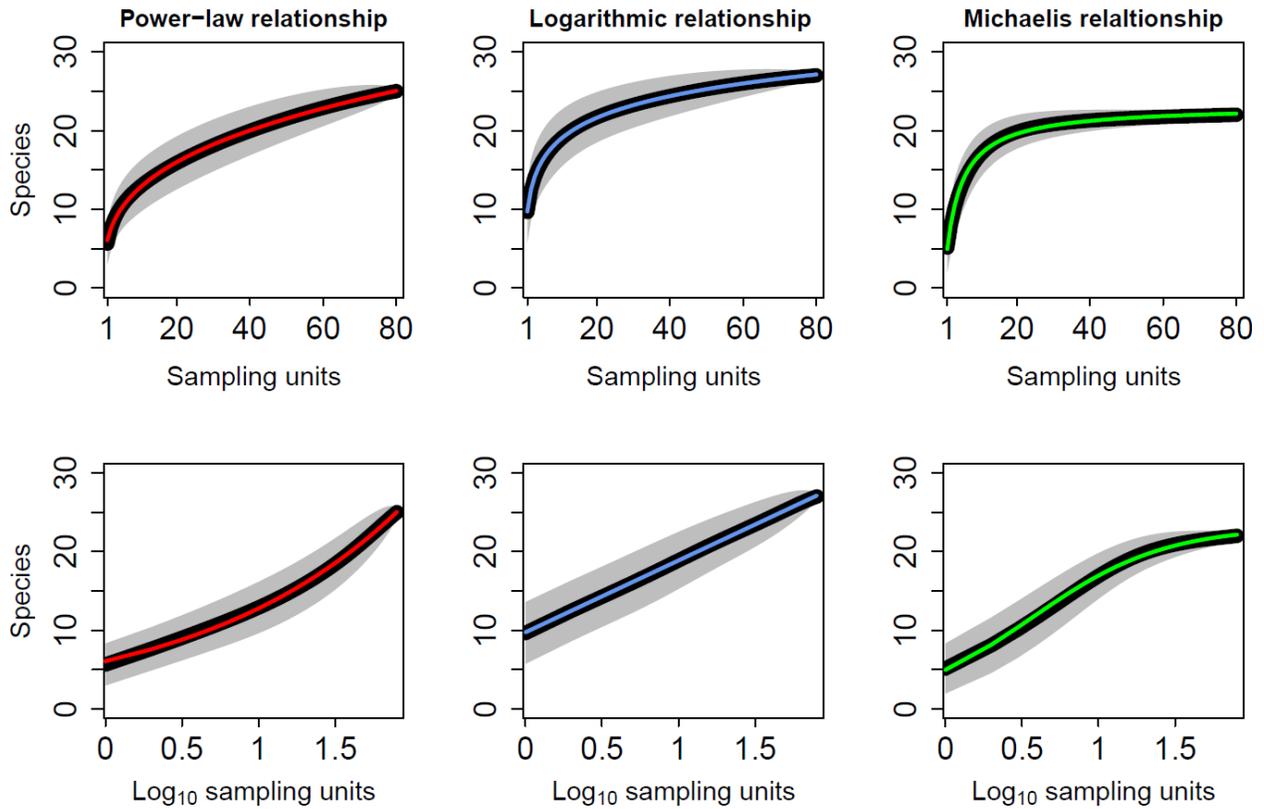

**Figure 2.**

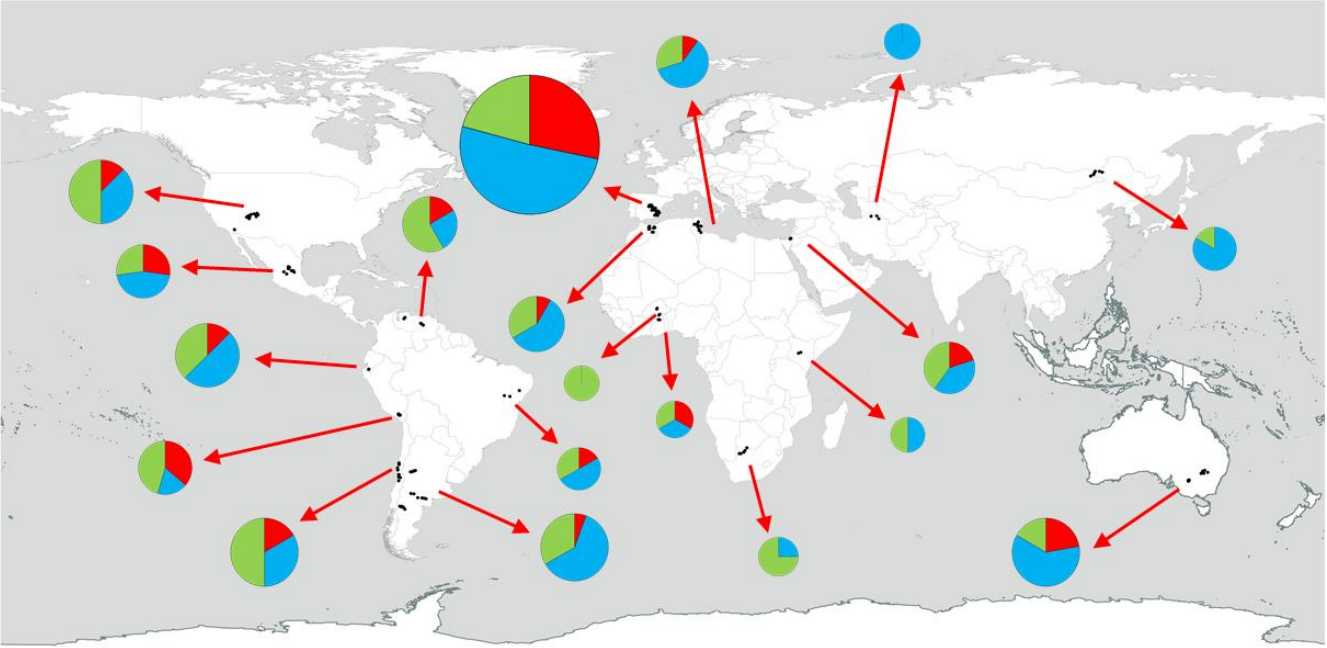

**Figure 3.**

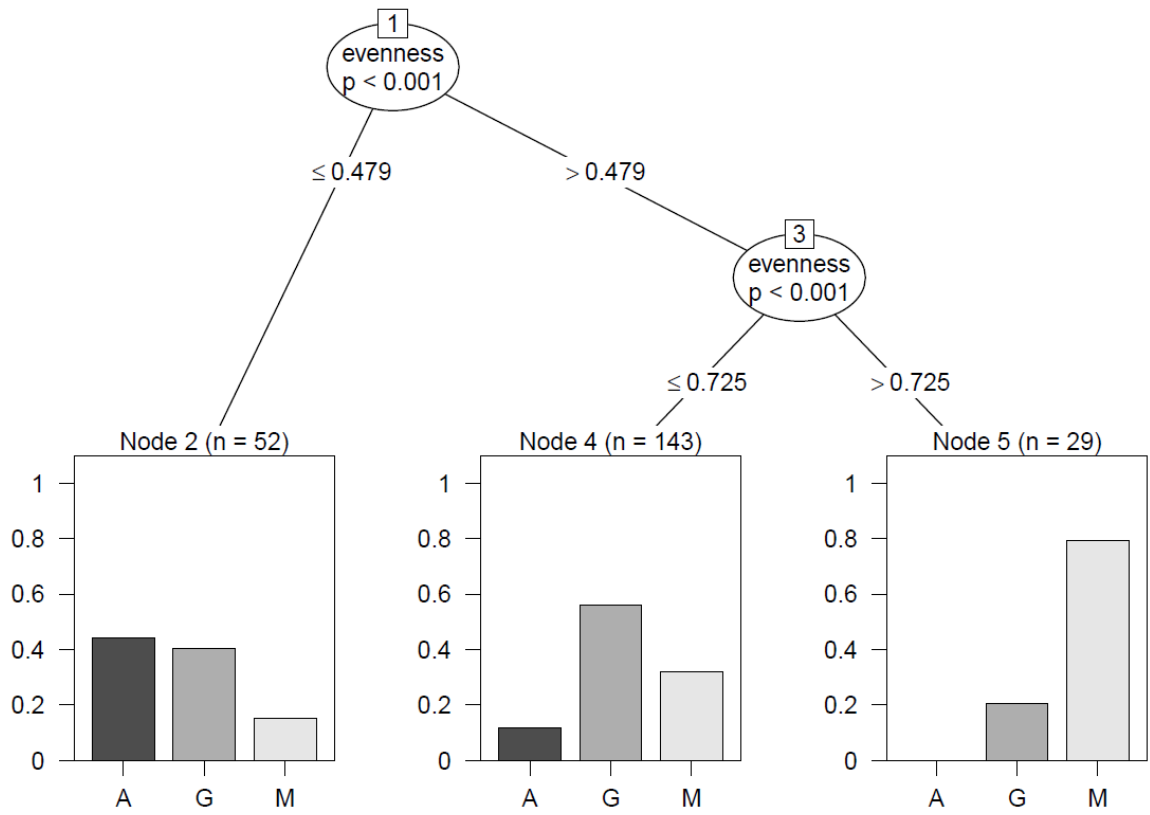

**Figure 4.**

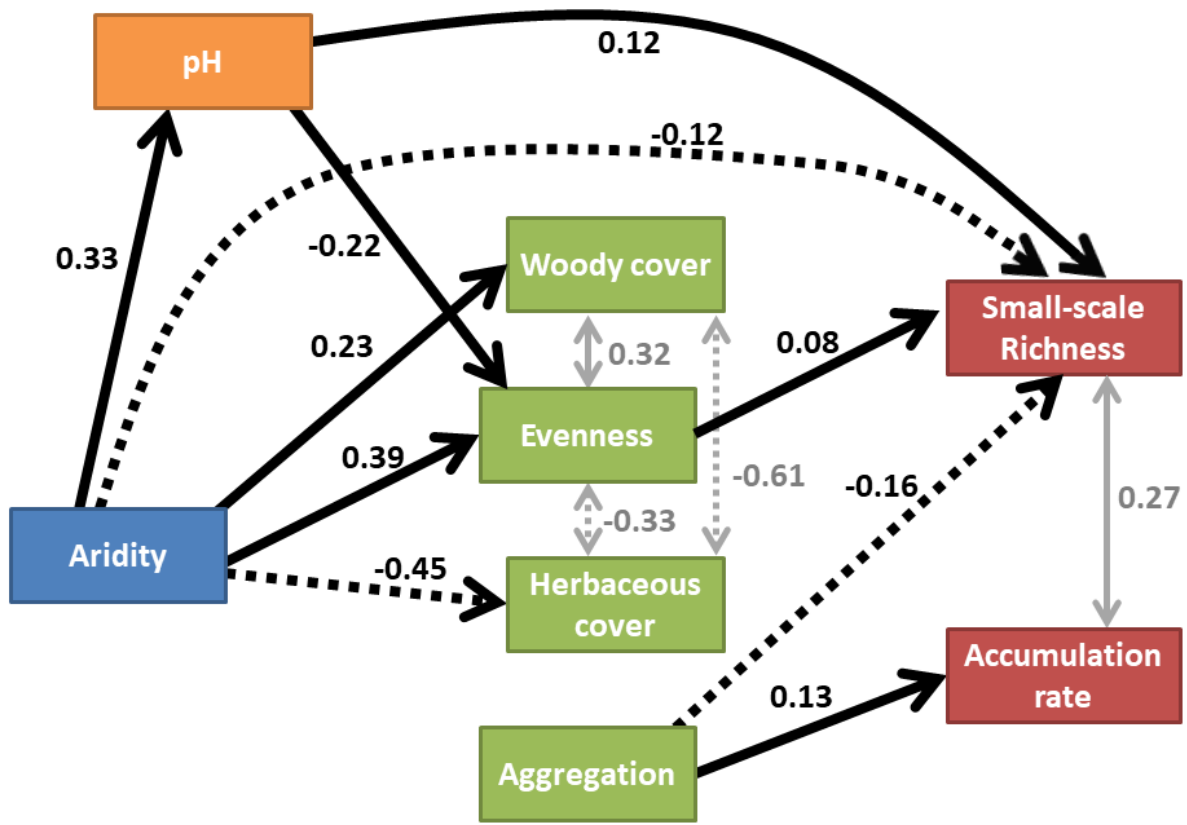

# SUPPLEMENTARY MATERIAL

## The structural equation model

The model includes the seven following equations:

(1) $pH = \alpha_{11} + \alpha_{12}\,Aridity + \alpha_{13}\,Spatial\ eigenvectors + e_1$

(2) $Woody\_Cover = \alpha_{21} + \alpha_{22}\,Aridity + \alpha_{23}\,pH + spatial\ eigenvectors + e_2$

(3) $Evenness = \alpha_{31} + \alpha_{32}\,Aridity + \alpha_{33}\,pH + \alpha_{34}\,Spatial\ eigenvectors + e_3$

(4) $Herbaceous\_Cover = \alpha_{41} + \alpha_{42}\,Aridity + \alpha_{43}\,pH + \alpha_{44}\,Spatial\ eigenvectors + e_4$

(5) $Aggregation = \alpha_{51} + \alpha_{52}\,Aridity + \alpha_{53}\,pH + \alpha_{55}\,Spatial\ eigenvectors + e_5$

(6) $Small\_Scale\_Richness = \alpha_{61} + \alpha_{62}\,Woody\_Cover + \alpha_{63}\,Evenness + \alpha_{64}\,Herbaceous\_Cover + \alpha_{65}\,Aggregation + \alpha_{66}\,Aridity + \alpha_{67}\,pH + \alpha_{68}\,Spatial\ eigenvectors + e_6$

(7) $Accumulation\_Rate = \alpha_{71} + \alpha_{72}\,Woody\_Cover + \alpha_{73}\,Evenness + \alpha_{74}\,Herbaceous\_Cover + \alpha_{75}\,Aggregation + \alpha_{76}\,Aridity + \alpha_{77}\,pH + \alpha_{78}\,Spatial\ eigenvectors + e_7$

Equation 1 describes the effect of *Aridity* on *pH*. Equations 2-5 describe the effects of *Aridity* and *pH* on the following variables: *Woody_cover, Evenness, Herbaceous_Cover* and *Aggregation* (hereafter 'mediators'). Equations 6-7 describe the effects *Aridity* and *pH* and the mediators on *Local_Scale_Richness* and *Accumulation_Rate* (the intercept and the slope of the species accumulation curve assuming logarithmic relationship). The notions $e_1$-$e_7$ represent the error terms.

*Aridity* and *pH* are also included in equations 6-7 for estimating effects of aridity that are not mediated by the potential 'mediators'. In addition, the model allows for non-causal correlation between the following variables: '*woody cover ~ herbaceous cover* ($r = -0.62$, $P < 0.001$), *Accumulation_Rate ~ Small-Scale Richness* ($r = 0.27$, $P < 0.001$), *Evenness ~ Herbaceous_cover* ($r = -0.33$, $P < 0.001$), *Woody_Cover ~ Evenness* ($r = 0.32$, $P < 0.001$), *Woody_cover ~ Aggregation* ($r = -0.08$, $P = 0.22$). The model was tested using a local estimation method (i.e. using a separate linear regression for each equation and later a combined test for conditional independence among the residuals).

Importantly, a robust analysis requires minimizing the problems of spatial autocorrelation (independence among nearby sites) that could affect the results (due to confounding factors). Hence, we used Moran Eigenvectors Maps that were built with the R package 'adespatial' [see 38]. These eigenvector describe the spatial pattern of the sites. Inclusion of these vectors in SEM equations enables the reduction of potential bias in parameter estimation caused by unmeasured factors related to spatial autocorrelation such as disturbances, historical land-use or soil characteristics. For reducing those confounding effects as much as possible (i.e. applying the most conservative approach), we included all the 38 positive eigenvectors in all SEM equations. in the SEM..

The model results are shown in table S1. In addition, bivariate scatter plots and the distribution of all variables (after the transformations) are shown in Fig S1.

## SUPPLEMENTARY TABLES

**Table S1:** Unstandardized and standardized coefficients and P-values obtained from the structural equation model

| Equation | Response | Predictor | Raw estimate | Standardized estimate | Standard error | P-value |
|---|---|---|---|---|---|---|
| 1 | pH | (logit) Aridity index | 0.423 | 0.325 | 0.090 | **<0.001** |
| 2 | (logit) woody cover | (logit) Aridity index | 0.331 | 0.229 | 0.110 | **0.003** |
| 2 | (logit) woody cover | pH | -0.147 | -0.132 | 0.086 | 0.090 |
| 3 | (logit) evenness | (logit) Aridity index | 0.436 | 0.387 | 0.080 | **<0.001** |
| 3 | (logit) evenness | pH | -0.192 | -0.222 | 0.062 | **0.002** |
| 4 | (logit) herbaceous cover | (logit) Aridity index | -1.153 | -0.449 | 0.182 | **<0.001** |
| 4 | (logit) herbaceous cover | pH | 0.207 | 0.105 | 0.142 | 0.148 |
| 5 | (log) aggregation | (logit) Aridity index | -0.005 | -0.023 | 0.016 | 0.755 |
| 5 | (log) aggregation | pH | -0.014 | -0.082 | 0.013 | 0.272 |
| 6 | (log) small-scale richness | (logit) woody cover | -0.112 | -0.044 | 0.110 | 0.311 |
| 6 | (log) small-scale richness | (logit) evenness | -0.251 | 0.078 | 0.127 | **0.049** |
| 6 | (log) small-scale richness | (logit) herbaceous cover | 0.028 | 0.020 | 0.066 | 0.672 |
| 6 | (log) small-scale richness | (log) aggregation | -2.667 | -0.161 | 0.592 | **<0.001** |
| 6 | (log) small-scale richness | (logit) Aridity index | -0.434 | -0.119 | 0.143 | **<0.001** |
| 6 | (log) small-scale richness | pH | 0.324 | 0.116 | 0.100 | **0.001** |
| 7 | (log) accumulation rate | (logit) woody cover | 0.006 | 0.009 | 0.032 | 0.860 |
| 7 | (log) accumulation rate | (logit) evenness | -0.055 | -0.071 | 0.037 | 0.138 |
| 7 | (log) accumulation rate | (logit) herbaceous cover | -0.028 | -0.084 | 0.019 | 0.141 |
| 7 | (log) accumulation rate | (log) aggregation | 0.494 | 0.126 | 0.171 | **0.004** |
| 7 | (log) accumulation rate | (logit) Aridity index | 0.068 | 0.078 | 0.041 | 0.103 |
| 7 | (log) accumulation rate | pH | -0.048 | -0.072 | 0.029 | 0.097 |

**SUPPLEMENTARY FIGURES**

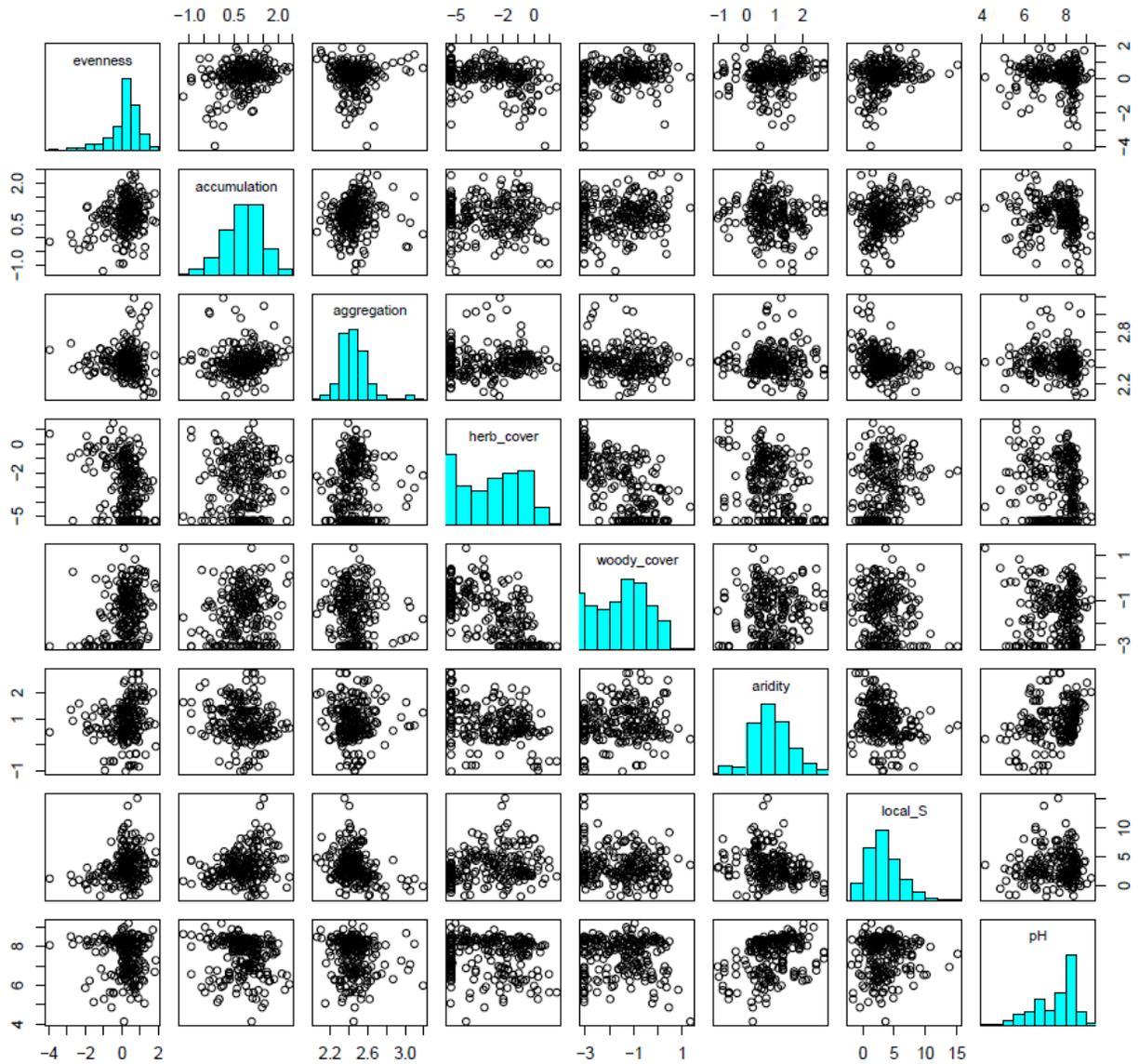

**Fig. S1:** Bivariate scatter plots and distribution of the variables used in the structural equation model (after transformations): evenness, accumulation rate (slope of the species accumulation curve), aggregation, herbaceous cover, woody cover, aridity index, local scale species richness and pH.